\documentclass[12pt]{article} 
\usepackage[xdvi]{graphicx} 
\usepackage{amssymb}
\usepackage{amsmath} 
 
\begin{document}   

\newcommand{\todo}[1]{{\em \small {#1}}\marginpar{$\Longleftarrow$}}   
\newcommand{\labell}[1]{\label{#1}\qquad_{#1}} 

\rightline{DCPT-02/23}   
\rightline{hep-th/0204209}   
\vskip 1cm


\begin{center} 
{\Large \bf Giant gravitons in non-supersymmetric backgrounds}
\end{center} 
\vskip 1cm   
  
\renewcommand{\thefootnote}{\fnsymbol{footnote}}   
\centerline{\bf   
 David C. Page\footnote{d.c.page@durham.ac.uk} and Douglas J. Smith\footnote{douglas.smith@durham.ac.uk}}    
\vskip .5cm   
\centerline{ \it Centre for Particle Theory, Department of  
Mathematical Sciences}   
\centerline{\it University of Durham, South Road, Durham DH1 3LE, U.K.}   
  
\setcounter{footnote}{0}   
\renewcommand{\thefootnote}{\arabic{footnote}}


\begin{abstract}   

We consider giant gravitons as probes of a class of ten-dimensional solutions of type IIB supergravity which arise as lifts of solutions of $U(1)^3$ gauged $\mathcal{N}=2$ supergravity in five-dimensions. Surprisingly it is possible to solve exactly for minimum energy configurations of these spherical D3-brane probes in the compact directions, even in backgrounds which preserve no supersymmetry. The branes behave as massive charged particles in the five non-compact dimensions. As an example we probe geometries which are believed to represent the supergravity background of coherent states of giant gravitons. We comment on the apparently repulsive nature of the naked singularities in these geometries.

\end{abstract}  

\section{Introduction}     

Giant gravitons are massless particles in supergravity which expand into branes. In $AdS \times S$
backgrounds they are massless particles with large angular momentum on the sphere.
As first described by McGreevy, Susskind and Toumbas~\cite{McGreevy:2000cw} these particles blow-up into branes which are expanded within the sphere.

A particularly interesting feature, shown by
Das, Trivedi and Vaidya~\cite{Das:2000ab} is that giant gravitons exist in
more general backgrounds including the near-horizon limits of non-extremal brane backgrounds which
preserve no supersymmetry.

Myers and Tafjord~\cite{Myers:2001aq} found evidence that particular type IIB
supergravity backgrounds (superstar geometries) contain giant gravitons, by
looking at dipole moments
of the five-form field strength, which are non-vanishing in the presence of expanded D3-brane sources. This was further investigated in the
analogous eleven-dimensional supergravity backgrounds~\cite{Balasubramanian:2001dx, Leblond:2001gn}. In~\cite{Leblond:2001gn}, giant graviton probe calculations were performed in the superstar background to show that in a simple case (with only one non-zero angular momentum on the sphere) the giant gravitons which were supposed to source the singularity of the geometry could consistently be placed at this singularity.

The ten-dimensional geometries investigated in~\cite{Myers:2001aq} arise as
lifts~\cite{Cvetic:1999xp} of certain gauged supergravity solutions describing
the extremal limit of charged black-holes in five
dimensions~\cite{Behrndt:1998ns, Behrndt:1998jd}. In trying to understand the
general multi-charge superstar geometry probe calculation which was left undone
in~\cite{Leblond:2001gn} we found that these calculations can be performed in
{\it any} background which is a lift of a solution of one of these gauged
supergravities. This shows that giant graviton states which are degenerate with massless particle states exist (at the level of classical solutions) in a large class of backgrounds which in general preserve no supersymmetry. This requires a number of remarkable cancellations between terms in the probe calculations. We comment on the possible relevance of this result in the conclusion.

The paper is structured as follows. We focus on the lifts of U$(1)^3$
gauged five-dimensional $\mathcal{N} = 2$ supergravity to type IIB supergravity
although similar results can be expected to hold for the closely related cases
of lifts of U$(1)^4$ four-dimensional and U$(1)^2$ seven-dimensional gauged
supergravities to
eleven-dimensional supergravity. In section 2 we review the truncation ansatz
and rearrange it into a suitable form for our calculations. In particular we
present the results of integrating the five-form field strength to find the
four-form gauge potential which couples to D3-branes. In section 3 we perform
a probe computation with a massless particle and see that it behaves like a
charged massive particle in five dimensions. We then show in section 4 that we
get the same result from a giant graviton probe computation. In section 5 we apply these results to the specific case of the ten-dimensional superstar geometries and show that in general the naked singularities of these spacetimes appear to repel the giant gravitons which are supposed to source them. We finish with some comments and suggestions for future investigation.

\section{The truncation ansatz}

The  bosonic fields of the five-dimensional $\mathcal{N} = 2$ supergravity with gauge group $U(1)^3$ are the metric, three $U(1)$ gauge fields $A^i$ (i=1,2,3) and two scalar fields which are usefully parametrized as $X_i$  obeying $X_1 X_2 X_3 = 1$.

The Lagrangian for the theory is 
\begin{equation}
  \label{eq:5dlagrangian}
  e^{-1} \mathcal{L} = R - \frac{1}{2} \sum_i (X_i^{-1} \partial X_i)^2 + \frac{4}{L^2} \sum_i X_i^{-1} - \frac{1}{4} \sum_i X_i^{-2}(F^i_{(2)})^2 - \frac{1}{4} e^{-1} \epsilon^{\mu \nu \rho \sigma \lambda} F^1_{\mu \nu} F^2_{ \rho \sigma} A^3_{\lambda}
\end{equation}

which leads to the equations of motion:
\begin{eqnarray}
  \label{eq:5deom1}
  d(X_1^{-2} *_{(1,4)} F^1_{(2)})  &=& - F^2_{(2)} \wedge F^3_{(2)} \rm{\qquad and \, cyclic \, permutations,}\\
  \label{eq:5deom2}
 d(X_i^{-1} *_{(1,4)} d X_i) &=&  \sum_j M_{i j} [\frac{4}{L^2} \epsilon_{(1,4)} X_j^{-1} - X_j^{-2} F^j_{(2)} \wedge *_{(1,4)} F^j_{(2)}] ,
\end{eqnarray}
where $ M_{i j} = \delta_{i j} - 1/3$.

The ten dimensional lift ansatz is \cite{Cvetic:1999xp}
\footnote{The lift in the case without the scalar fields (i.e.\ $X_i = 1$) was
first found in \cite{Chamblin:1999tk}.}:
\begin{equation}
\label{metricansatz}
ds_{10}^2 = \Delta^{1/2} ds_{1,4}^2   + \Delta^{-1/2} \sum_i X_i^{-1} (L^2 d \mu_i ^2 + \mu_i ^2[L d \phi_i + A^i]^2 ) 
\end{equation}
and $F_{(5)} = G_{(5)} + *G_{(5)}$ where
\begin{eqnarray}
 G_{(5)} &=& \frac{2}{L}  \sum_i (X_i^{2}  \mu_i ^2 -  \Delta X_i) \epsilon_{(1,4)} + \frac{L}{2} \sum_i  X_i^{-1} *_{(1,4)} d X_i \wedge d(\mu_i^{2}) \nonumber \\
& +& \frac{L}{2}  \sum_i  X_i^{-2}  d(\mu_i^{2}) \wedge [L d\phi_i + A^i] \wedge *_{(1,4)} F^i_{(2)}
\label{5form}
\end{eqnarray}

Here, $ds_{1,4}^2$ and $ \epsilon_{(1,4)}$ are the metric and volume form of the five dimensional geometry and $*_{(1,4)}$ denotes the Hodge dual with respect to the five dimensional metric.  $\Delta$ is given by 
\begin{equation}
\Delta =  \sum_i X_i \mu_i ^2 .
\end{equation}
The three quantities $\mu_i$ are non-negative real variables subject to the
constraint $\sum_i \mu_i ^2 = 1$ and we  parametrize them by
\begin{equation}
\mu_1 = \cos \theta_1 , \, \, \mu_2 = \sin \theta_1 \cos \theta_2 , \, \, \mu_3 = \sin \theta_1 \sin \theta_2 .
\label{mu_theta}
\end{equation}
where $0 \le \theta_1 \le \pi/2$ and $0 \le \theta_2 \le \pi/2$. The angles
$\phi_i$ are unconstrained with $0 \le \phi_i < 2\pi$. I.e.\ each pair
$\{\mu_i, \phi_i\}$ are planar polar coordinates and the constraint on the
$\mu_i$ describes the embedding of a unit $S^5$ in flat $\mathbf{R}^6$.

In what follows, we will need explicit expressions for $F_{(5)}$ and for the potential $A_{(4)}$ satisfying $dA_{(4)} = F_{(5)}$ which gives the coupling to D3-brane probes. The calculations needed to find these expressions can be found in appendices \ref{GtoStarG} and \ref{IntF5}, but the final answers are rather simple. The two pieces in the self-dual five-form $F_{(5)}$ are $G_{(5)}$, given above in equation~(\ref{5form}) and 
\begin{eqnarray}
*G_{(5)} &=& -\frac{2}{L \Delta^2}  \sum_i (X_i^{2}  \mu_i ^2 - \Delta X_i) L^2 W  \bigwedge_k \mu_k [L d \phi_k + A^k] \nonumber \\
   & &  - \sum_i \partial_{\nu} \left( \frac{X_i \mu_i}{\Delta} \right) dx^{\nu} \wedge L Z_i \bigwedge_k \mu_k [L d \phi_k + A^k] \nonumber \\
  & & + \frac{L}{\Delta} \sum_{i j} X_i \mu_i Z_{ij} \bigwedge_{k \neq j} \mu_k [L d \phi_k + A^k] \wedge F^j_{(2)}
\label{StarG5}
\end{eqnarray}
where $W =  \sin \theta_1 d\theta_1 \wedge d\theta_2$ is the volume form on the
two-sphere spanned by $\mu_1, \mu_2, \mu_3$ and where
\footnote{We use similar notation to \cite{Cvetic:2000eb}.}
$Z_i = \epsilon_{ijk} \mu^j d\mu^k$, $Z_{ij} = \epsilon_{ijk} d\mu^k$. For
one-forms $d\phi_i$ our notation is:
$$ \bigwedge_{k \neq i} d\phi_k = \frac{1}{2} \sum_{jk} \epsilon_{ijk} d\phi_j \wedge d\phi_k $$

As shown in appendix B, $F_{(5)}$ can (almost) be written in a manifestly closed form
from which it is easy to read off the pieces of $A_{(4)}$:
\begin{eqnarray}
 A_{(4)} & = & \frac{L}{2} \sum_i \mu_i^2 \left[ X_i^{-1} *_{(1,4)} d X_i + X_i^{-2} [Ld\phi_i + A^i] \wedge  *_{(1,4)} F^i_{(2)} \right] \nonumber \\
 & & -\frac{L}{2}\mu_2^2 \bigwedge_{k \ne 3} [L d \phi_k + A^k] \wedge F^3_{(2)} +\frac{L}{2}\mu_3^2 \bigwedge_{k \ne 2} [L d \phi_k + A^k] \wedge F^2_{(2)} \nonumber \\
 & & +\frac{L}{\Delta \mu_1}\sum_i X_i \mu_i Z_{1i} \bigwedge_k \mu_k [L d \phi_k + A^k] + \tilde{A}_{(4)}
\label{eq:A4}
\end{eqnarray}
where $\tilde{A}_{(4)}$ is a four-form satisfying
\begin{equation}
d \tilde{A}_{(4)} = - \frac{4}{3L} \sum_i X_i^{-1} \epsilon_{(1,4)}
	- \frac{L}{6}\sum_i X_i^{-2} F^i_{(2)} \wedge  *_{(1,4)} F^i_{(2)}
	- \frac{L}{2} [Ld\phi_1 + A^1] \wedge F^2_{(2)} \wedge F^3_{(2)}
\end{equation}
The precise form of the piece of $\tilde{A}_{(4)}$ in the five-dimensional
spacetime will depend on the particular solution for the $X_i$ and is not
important for the calculations we will perform. Note that $A_{(4)}$ is only
locally well-defined since it becomes singular if $\mu_1$ vanishes
\footnote{There are of course similar alternative expressions which become
singular if $\mu_2$ or $\mu_3$ vanish. Note that these singularities are
unavoidable since e.g.\ $F_{(5)}$ contains the volume-form
on the 5-sphere which is closed but not exact.}.
In section \ref{GGProbes} we will consider a D3-brane at constant
$\theta_1$ so the above expression is well-defined
(for $ \mu_1 = \cos \theta_1 \ne 0$) and, for example, we can read off, after
substituting equations~(\ref{mu_theta}) and some simple manipulations,
\begin{equation}
A^{(4)}_{\phi_1\theta_2\phi_2\phi_3} =
- \frac{L^4}{\Delta} \sin^4 \theta_1 \cos \theta_2 \sin \theta_2 
\left( X_2 \cos ^2 \theta_2 +  X_3 \sin^2 \theta_2 \right)
\end{equation}

\section{A massless particle probe}

As a warm up for what follows we consider the action of a massless particle which carries some conserved angular momentum on the five-sphere but otherwise minimizes its energy in the internal space. For example a massless particle which is stationary on the internal sphere would appear simply as a massless particle in the remaining five dimensions. More generally we might expect that a massless particle with some angular momentum would appear massive and charged in five-dimensions.

For convenience we start from the action for a massive particle in
ten-dimensions and later take the mass $m$ to zero.
\begin{equation}
  \label{eq:particleprobe}
  S = -m \int dt \sqrt{-\det(\mathcal{P}(g))}  
\end{equation}
where $\mathcal{P}(g)$ is the pullback of the spacetime metric onto the particle's world line and is given by
\begin{equation}
  \label{eq:pullback}
  \mathcal{P}(g) = g_{m n} \dot{x}^m \dot{x}^n
\end{equation}
Here $x^{m}$ are coordinates on ten-dimensional space with $x^0 = t$ and
$\dot{x}^m$ is the derivative of $x^m$ wrt.\ $t$. The metric $g_{m n}$ is given by equation~(\ref{metricansatz}) and it is convenient to use coordinates $x^{\mu} (\mu = 0 \ldots 4), \theta_1, \theta_2$ and $ \phi_i$. We consider a particle with angular momenta in the $ \phi_i$ directions but stationary in the $\theta_1, \theta_2$ directions. The Lagrangian becomes:
\begin{equation}
  \label{eq:particleprobe2}
  \mathcal{L} = -m  \sqrt{- \Delta^{1/2}(g_{\mu \nu}\dot{x^{\mu}}\dot{ x^{\nu}}) - \Delta^{-1/2} \sum_i X_i^{-1} \mu_i^2 (L \dot{\phi_i} +  A^i_{\mu} \dot{x^{\mu}})^2}
\end{equation}
Since all the fields in the lift ansatz are independent of $\phi_i$, the action we have written down has no explicit $\phi_i$ dependence and we can replace $\dot{\phi_i}(t)$ with conjugate momenta $P_i$ which are conserved in time. The resulting Routhian is found to be:

\begin{equation}
\mathcal{R} = \sqrt{-g_{\mu \nu}\dot{x^{\mu}}\dot{ x^{\nu}}} \left( \sum_i \frac{P_i^2 X_i \Delta}{L^2\mu_i^2}\right)^{1/2} - \frac{1}{L} \sum_i \dot{x^{\mu}} A^i_{\mu} P_i.
\end{equation}
where we have now taken the limit $m \rightarrow 0$.

We need to find the minimum of the energy with respect to the $\mu_i$. 
If we define vectors $\mathbf{U}$ and $\mathbf{V}$ by:
\begin{eqnarray*}
  U_i &=& \sqrt\frac{X_i}{\mu_i^2} \frac{P_i}{L} \, \mbox{  for i = 1,2,3} \\
V_i &=&  \sqrt{X_i \mu_i^2}  \, \mbox{  for i = 1,2,3} 
\end{eqnarray*}
Then 
\begin{eqnarray*}
  \left( \sum_i \frac{P_i^2 X_i \Delta}{L^2\mu_i^2}\right)^{1/2} &=& |\mathbf{U}| |\mathbf{V}|  \\
&\geq&  \mathbf{U}.\mathbf{V}  \qquad  \mbox{  (Schwarz inequality)}
\end{eqnarray*}
with equality iff.\ $\mathbf{U}$ and $\mathbf{V}$ are parallel. Thus the minimum of the energy occurs at $\mu_i^2 = (P_i / \sum_j P_j)$, taking into account the constraint $\sum_i \mu_i^2 = 1$.

The resulting charged particle Lagrangian in five dimensions is:

\begin{equation}
  \label{eq:Chargedparticle}
  \mathcal{L} = \frac{1}{L} \sum_i \left(-P_i X_i\sqrt{-g_{\mu \nu}\dot{x^{\mu}}\dot{ x^{\nu}}}  + \dot{x^{\nu}} A^i_{\nu} P_i\right).
\end{equation}

It should be noted that in the calculation above, the choice of
five-dimensional geometry (and background fields) and the motion of the
particle probe in the five non-compact directions remains arbitrary throughout.
It is perhaps rather surprising that it is possible to minimize the energy of a
massless probe in the compact space independently of these details. Presumably
this is illustrating some special properties of the lift ansatz. It would be
interesting to understand this in more detail.

\section{Giant graviton probes}
\label{GGProbes}

In this section we consider probing the ten-dimensional lift ansatz of section 2 with giant gravitons. We find that, as with the massless particle probes of the previous section, it is possible to find minimum energy configurations in the compact directions without specifying a particular five-dimensional solution or any particular motion of the probe in the five non-compact dimensions. As with the massless particle probes, the giant gravitons behave simply as massive charged particles in five dimensions.

Our giant graviton probe will be a D3-brane with the topology of an $S^3$ lying inside the $S^5$. In particular the brane
will `wrap' the $\theta_2$, $\phi_2$ and $\phi_3$ directions while moving
rigidly in the $\phi_1$ direction at fixed $\theta_1$.
The motion of the probe in the non-compact directions remains arbitrary with
the assumption that it is independent of $\theta_2$, $\phi_2$ and $\phi_3$, i.e.\ we only consider rigid motion of the brane. While it is not
initially obvious that this is a consistent way of embedding the brane, we will
see that it does in fact give a minimal energy configuration. Specifically we
find that the brane action reduces to a particle action in five dimensions --
independent of $\theta_2$, $\phi_2$ and $\phi_3$ and with $\theta_1$ and
$\dot{\phi}_1$ constant. 

The action for the D3-brane probe is:

\begin{equation}
  S_3  =  -T_3 \int dt d\theta_2 d\phi_2 d\phi_3 \left[  \sqrt{-\det(\mathcal{P}(g))}           + \dot{x^{\nu}} A^{(4)}_{\nu \theta_2 \phi_2 \phi_3}  +  \dot{\phi_1} A^{(4)}_{\phi_1 \theta_2 \phi_2 \phi_3}\right]
 \label{eq: Probeaction}
\end{equation}

Our first task in evaluating this action will be to find the pieces of the RR four-form potential $A^{(4)}$ which couple to the probe.
It is straightforward to read off the relevant pieces of $A^{(4)}$ from equation~(\ref{eq:A4}). We find 
\begin{equation}
   \dot{x^{\nu}} A^{(4)}_{\nu \theta_2 \phi_2 \phi_3}  +  \dot{\phi_1} A^{(4)}_{\phi_1 \theta_2 \phi_2 \phi_3} = - \frac{L^3}{\Delta} \sin^4 \theta_1 \cos \theta_2 \sin \theta_2 \alpha \dot{\Phi}
\end{equation}
where $\alpha =  X_2 \cos ^2 \theta_2 +  X_3 \sin^2 \theta_2$ and
$\dot{{\Phi}} \equiv L\dot{\phi_1} + \dot{x^{\nu}} A_{\nu}^1$.

Combining this with the terms coming from the pullback of the metric we find

\begin{eqnarray}
  S_3  &=&  -T_3 L^3 \int dt d\theta_2 d\phi_2 d\phi_3 \left[ \frac{\sin^3 \theta_1}{\sqrt{\Delta}} X_1 \cos\theta_2 \sin\theta_2 \alpha^{1/2} \sqrt{-g_{\mu\nu}\dot{x}^{\mu}\dot{x}^{\nu} - \frac{\cos^2\theta_1}{X_1 \Delta}\dot{\Phi}^2} \right. \nonumber \\
 & &  \left.  - \frac{1}{\Delta} \sin^4 \theta_1 \cos \theta_2 \sin \theta_2 \alpha \dot{\Phi} \right]
 \label{eq: Probeaction2}
\end{eqnarray}

Since all the fields in the lift ansatz are independent of $\phi_1$, the action
we have written down has no explicit $\phi_1$ dependence and we can replace
$\dot{\phi_1}$ with a conjugate momentum $P_{\phi_1}(\theta_2,\phi_2,\phi_3)$
which is independent of time. The resulting Routhian is found to be:

\begin{eqnarray}
  \mathcal{R} &=& \frac{1}{L} \sqrt{-g_{\mu\nu}\dot{x}^{\mu}\dot{x}^{\nu}} \sqrt{ \frac{ X_1 \Delta}{\cos^2 \theta_1} \left(P_{\phi_1} - N\frac{ \alpha \sin^4 \theta_1 \cos \theta_2 \sin \theta_2}{\Delta V_3} \right)^2 + N^2\frac{ \sin^6 \theta_1 X_1^2 \alpha \cos^2 \theta_2 \sin^2 \theta_2}{\Delta V_3^2}} \nonumber \\
& & \qquad \qquad - \frac{1}{L} \dot{x^{\nu}} A^1_{\nu} P_{\phi_1} \nonumber \\
&=& \frac{1}{L} \sqrt{-g_{\mu\nu}\dot{x}^{\mu}\dot{x}^{\nu}} X_1 \sqrt{ P_{\phi_1}^2 + \frac{\alpha}{X_1} \tan^2 \theta_1 (P_{\phi_1} - \frac{1}{V_3} N\sin^2 \theta_1 \cos \theta_2 \sin \theta_2)^2 } - \frac{1}{L} \dot{x^{\nu}} A^1_{\nu} P_{\phi_1},
\label{Routhian}
\end{eqnarray}
where $N = V_3 T_3 L^4$ and $V_3 = 2\pi^2$ is the volume of a unit $S^3$ in
${\mathbf R}^4$.

Minimizing over $\theta_1$ is now straightforward. The expanded minimum occurs
when $P_{\phi_1}(\theta_2,\phi_2,\phi_3) = \frac{1}{V_3} P_1 \cos\theta_2 \sin \theta_2$
for some constant $P_1$ and $N\sin^2 \theta_1 = P_1$. 
(There is also a zero-size minimum when $\theta_1 = 0$ corresponding to the massless particle probe of the previous section.) It is easy to check that this form for $P_{\phi_1}$ implies that $\dot{\phi_1}$ is constant. If we substitute these values back into the Routhian and integrate over $\theta_2$, $\phi_2$ and $\phi_3$ the resulting particle Lagrangian is:

\begin{equation}
  \label{eq:Chargedparticle2}
  \mathcal{L} = \frac{1}{L} \left( -P_1 X_1\sqrt{-g_{\mu\nu}\dot{x}^{\mu}\dot{x}^{\nu}}  + \dot{x^{\nu}} A^1_{\nu} P_1 \right)
\end{equation}

It is remarkable that the complicated action of equation~(\ref{eq: Probeaction2}) should reduce to such a simple form when we consider minimum energy configurations in the compact directions. The crucial step is that the Routhian~(\ref{Routhian}) rearranges into a sum of squares to make minimization over $\theta_1$ simple. That this occurs in the absence of supersymmetry is somewhat surprising and merits further investigation.

Equation~(\ref{eq:Chargedparticle2}) is the same Lagrangian for a charged particle in five dimensions which we saw in the previous section. The conclusion is that probing with the giant graviton is equivalent to probing with a massless particle in ten-dimensions and both are equivalent to probing with a charged massive particle in five-dimensions.

\section{Superstars and giant gravitons}

In this section we apply the giant graviton probe results which we have found for general lifts of the five-dimensional supergravity, to study a specific example - the superstar geometries of~\cite{Cvetic:1999xp, Myers:2001aq}. These geometries are believed to be the supergravity backgrounds corresponding to a collection of giant gravitons in $AdS_5 \times S^5$.
 
The five-dimensional $\mathcal{N}=2$ supergravity admits charged $AdS$ black
hole solutions~\cite{Behrndt:1998ns, Behrndt:1998jd} which in the extremal
limit can be written in the form
\begin{equation}
\label{chargedbh1}
ds^2_{1,4} = -(H_1 H_2 H_3)^{-2/3} f dt^{2} + (H_1 H_2 H_3)^{1/3}
(f^{-1} r^2 dr^{2} + d\Omega_3^2),
\end{equation}
\begin{equation}
\label{chargedbh2}
A_{(1)}^{i} = -q_i H_i^{-1} dt \qquad i=1,2,3,
\end{equation}
\begin{equation}
\label{chargedbh3}
X_{i} = H_i^{-1} (H_1 H_2 H_3)^{1/3},
\end{equation}
where we have introduced
\begin{equation}
f = (r^2 - q_3)^2 + \frac{1}{L^2}{H_1 H_2 H_3},
\end{equation}
\begin{equation}
H_i = r^2 + q_i - q_3,
\end{equation}
The $q_{i}$'s are the three $U(1)$ charges and without
loss of generality we have chosen $q_1 \ge q_2 \ge q_3 \ge 0$. In this extremal
limit there is a naked singularity at $r=0$.

These solutions can be lifted to ten-dimensions using the lift ansatz, equations~(\ref{metricansatz}) and (\ref{5form}), giving rise to the type IIB superstar geometries. We shall not use the explicit form of the ten-dimensional solution here (it can be found in references~\cite{Cvetic:1999xp, Myers:2001aq}), but we note that the five-form $F_{(5)}$ contains a piece:
\begin{eqnarray}
  \label{eq:dipole}
  F_{(5)} &=& \frac{L}{2}  \sum_i  X_i^{-2}  d(\mu_i^{2}) \wedge [L d\phi_i + A^i] \wedge *_{(1,4)} F^i_{(2)} = \ldots \nonumber \\
&=& L \sum_i q_i  d(\mu_i^{2}) \wedge  [L d\phi_i + A^i] \wedge e_{\alpha_1} \wedge e_{\alpha_2} \wedge e_{\alpha_3} + \ldots
\end{eqnarray}
where the $e_{\alpha_i}$ are a vielbein on the sphere in the AdS space with metric $d \Omega_3^2$. It was argued in~\cite{Myers:2001aq} that this corresponds to the dipole sourced by a collection of giant gravitons at $r=0$. If $n_i$ represents the total number of giant gravitons with non-zero momentum in the $\phi_i$ direction then the density of such giant gravitons can be shown to be~\cite{Myers:2001aq}:
\begin{equation}
  \frac{d n_i}{d \mu_i} = - 2 \frac{N q_i}{L^2} \mu_i.
\end{equation}
Using the result $P_1 = N \sin^2 \theta_1$, (or more generally $P_i = N(1 - \mu_i^2)$,) derived in the previous section, this shows that the total angular momentum of the geometry carried by each set of giant gravitons is:
\begin{equation}
  P^{total}_i = \frac{N^2}{2} \frac{q_i}{L^2}.
\end{equation}
This is equal~\cite{Myers:2001aq} to the total angular momentum carried by the five-dimensional geometry~\cite{Behrndt:1998jd}. The probe calculations of the previous section provide the necessary justification for the use of the formula  $P_i = N(1 - \mu_i^2)$ in the general superstar backgrounds.

In order to identify the naked singularities in the superstar geometries as corresponding to a collection of giant gravitons  we should further check whether giant graviton probes
minimize their energy at the singularity. We consider giant graviton probes
carrying angular momentum in the $\phi_i$ direction. The results of the
previous section show that it is equivalent to probe the five-dimensional
charged black holes with the charged particle probe of
equation~(\ref{eq:Chargedparticle2}). We insert the fields of
equations~(\ref{chargedbh1})--(\ref{chargedbh3}) into the probe
action~(\ref{eq:Chargedparticle2}) and look for a stationary solution
($\dot{x}^{\mu} = 0$ for $\mu \neq 0$.) The resulting energy of the probe is:
\begin{equation}
E_i = \frac{P_i}{L} \frac{f^{1/2} +q_i}{H_i}
\end{equation}

The question we are interesting in answering is whether the singularity $r=0$
corresponds to a BPS minimum ($E_i = P_i/L$) for each type of probe. It turns
out that there are several distinct cases to consider.
\begin{itemize}
\item
$q_1=q_2=q_3=0$ -- All three types of probe have a BPS minimum at $r=0$ as
expected.

\item
$q_2=q_3=0$ with $q_1$ non-zero -- The probe coupling to $A^1$ has a BPS
minimum while the probes coupling to $A^2$ or $A^3$ appear to have non-BPS
minima ($E_{2,3} = \sqrt{1+q_{2,3}/L^2}P_{2,3}/L$) at $r=0$.

\item
$q_3=0$ with $q_1$ and $q_2$ non-zero -- For the probes coupling to $A^1$ or
$A^2$, the energy saturates the BPS bound at $r=0$ but the gradient of the
potential is non-zero indicating an attractive force at the singularity. For
the probe coupling to $A^3$ the energy diverges as $r \rightarrow 0$ and there
is an infinite repulsive force. 

\item
All $q_i$ non-zero -- The singularity is repulsive to all three types of
probe. Furthermore there is an infinite repulsive force on the probe
coupling to $A^3$ (and on the probes coupling to $A^i$ in the special cases
$q_i = q_3$.)
\end{itemize}

The meaning of these results is not entirely clear. For a singly charged superstar, the fact that a probe with the same type of charge as the background has a BPS minimum energy configuration at the singularity agrees with the interpretation of the singularity as a collection of giant gravitons. However, in the other cases agreement is not reached. A possible resolution is that curvature corrections to the supergravity background and to the D3-brane action could modify the results in these less supersymmetric cases.
 
\section{Conclusions}

We have considered giant gravitons probing solutions of type IIB supergravity which are lifts of solutions of a five-dimensional gauged supergravity.  In
particular we have shown that the structure of the lift ansatz ensures that the
action for a giant graviton reduces to that of a massive charged particle in
five dimensions. The mass and charge of this particle are equal, suggesting that this is the
bosonic part of a superparticle action. So it seems that the consistent
truncation ansatz applies not only to the pure supergravity fields but also to
allowed sources in the form of brane actions which can be coupled to the
supergravity action. The derivation of the particle action in this way is
similar to the derivation of type IIA string and D-brane actions from M-brane
actions using the truncation of eleven-dimensional supergravity to
ten-dimensional type IIA supergravity \cite{Duff:1987bx, Townsend:1996af}.

In section 5 we looked at a specific example of a charged $AdS$ black hole
solution of five-dimensional gauged supergravity which lifts to the superstar
geometry in ten dimensions discussed in \cite{Myers:2001aq, Leblond:2001gn}.
We found, in the multi-charge case, that charged particle probes are repelled
by the naked singularity and hence from the results of section 4, giant
graviton probes are repelled by the superstar naked singularity. A possible
conclusion might be that these singular geometries are unstable or that they
are not sourced by giant gravitons. However, there is also a possibility that
higher curvature corrections might change the situation. Irrespective of the
precise interpretation, this is just a specific example of the more general
result from section 4 that we can investigate the nature of singularities (at
least within the context of giant graviton probes) without lifting to the
ten-dimensional solution.

Giant gravitons in $AdS \times S$ or $AdS_5 \times T^{1,1}$ spacetimes are BPS
branes preserving
half the background supersymmetry. In \cite{Mikhailov:2000ya} Mikhailov
presented an elegant construction of the preserved supercharges and in doing
so generalised the construction of giant gravitons to BPS objects preserving
one quarter or one eighth of the background supersymmetry. This description is
closely related to the amount of supersymmetry preserved by planar branes
or branes wrapped on holomorphic curves in Minkowski spacetime. It would be
interesting to examine these generalised giant gravitons in the context of
general gauged supergravity backgrounds to see whether the lift ansatz
provides the appropriate minimal surfaces for the branes to wrap even in the
absence of supersymmetry, as we have seen for the usual giant gravitons.

\vskip.5in

\centerline{\bf Acknowledgements}
\medskip    

We wish to thank
Ian Davies, James Gregory, Emily Hackett-Jones, Clifford Johnson, Ken Lovis, Rob Myers, Tony Padilla,
Chris Pope and Simon Ross
for helpful discussions and correspondence.
DCP is funded by the Engineering and Physical Sciences Research Council.

\appendix

\section{From $G_{(5)}$ to  $*G_{(5)}$}
\label{GtoStarG}

In order to dualise  $G_{(5)}$ we will need to dualise several forms in ten-dimensions which split into a p-form, $\alpha_{(p)}$, in the AdS directions and a q-form, $\beta_{(q)}$, in the sphere directions. The following result will be useful:
\begin{equation}
  *_{(10)} (\alpha \wedge \beta) = (-)^{q(5-p)} \Delta^{(q-p)/2} (*_{(1,4)} \alpha \wedge *_{(5)} \beta).
\end{equation}
The remaining difficulty in dualising $G_{(5)}$ resides in the fact that the sphere metric is given in terms of constrained variables (the $\mu_i$'s.)
Consider $R^3$ spanned by the  $\mu_i$'s (without the constraint $\sum_i \mu_i^2 = 1$) and with metric:
\begin{equation}
  ds^2_3 = \sum_i X_i^{-1} d \mu_i^2
\end{equation}
Let S be the surface given by $\sum_i \mu_i^2 = 1$ and denote the restriction of this metric to S by  $ds^2_2$  . Suppose that $e_1, e_2, e_3$ are a vielbein for  $R^3$ with the metric $ds^2_3$, such that $e_3 = \Delta^{-1/2} \sum_i \mu_i d \mu_i$. Then the following identity holds for dualising forms within the surface S:
\begin{equation}
  *_{(2)} \alpha = *_{(3)} (e_3 \wedge \alpha ).
\end{equation}
Thus, for example
\begin{eqnarray}
  Vol_S &=&  *_{(2)} 1 =  *_{(3)} \Delta^{-1/2} \sum_i \mu_i d \mu_i = \Delta^{1/2} W, \\
 \ast_{(2)} d(\mu_i^2) &=&  *_{(3)} \Delta^{-1/2} \sum_j \mu_j d \mu_j \wedge d(\mu_i^2) = -2  \Delta^{-1/2} \sum_j X_i X_j  \mu_i \mu_j Z_{i j}
\end{eqnarray}
where $W = \frac{1}{2} \sum_{ijk} \epsilon_{ijk} \mu_i d \mu_j \wedge d \mu_k$ is the volume form on the sphere $\sum_i \mu_i^2 = 1$ embedded in flat $R^3$ and $Z_{i j} = \sum_k \epsilon_{ijk} d \mu_k$.

It is now relatively straightforward to dualise $G_{(5)}$ as given by
equation~(\ref{5form}). We find:
\begin{eqnarray}
 \ast G_{(5)} = &-& L \frac{2U}{\Delta^2} W \bigwedge_i \mu_i [L d \phi_i + A^i] \nonumber \\
&+& -\frac{L}{\Delta^2} \sum_{i j} X_j dX_i \wedge Z_{i j} \mu_i \mu_j \bigwedge_l \mu_l [L d \phi_l + A^l] \nonumber\\
&+& -\frac{L}{\Delta} \sum_{i j} F^i_{(2)} \wedge Z_{i j} \mu_j X_j \bigwedge_{l \neq i} \mu_l [L d \phi_l + A^l] \nonumber
\end{eqnarray}
where $U = \sum_i(X_i^{-2}\mu_i^2 - \Delta X_i)$. This expression for $*G_{(5)}$
agrees (after some straightforward algebra) with equation~(\ref{StarG5}).

\section{Integrating $F_{(5)}$}
\label{IntF5}
Here we show that the forms of $F_{(5)} = dA_{(4)}$ where $A_{(4)}$ is given
in equation~(\ref{eq:A4}) and
of $G_{(5)} + *G_{(5)}$ given in equations~(\ref{5form}) and (\ref{StarG5}) are
equal when the five-dimensional equations of motion hold. We start by
evaluating
\begin{eqnarray}
  & & d \left( \frac{L}{2} \sum_i \mu_i^2 \left[ X_i^{-1} *_{(1,4)} d X_i +
	X_i^{-2} [Ld\phi_i + A^i] \wedge  *_{(1,4)} F^i_{(2)} \right] \right) \nonumber \\
&=&   \frac{L}{2} \sum_i d (\mu_i^2) \wedge X_i^{-1} *_{(1,4)} d X_i  + \frac{L}{2} \sum_i  \mu_i^2 d( X_i^{-1} *_{(1,4)} d X_i) \nonumber \\
& & + \frac{L}{2} \sum_i d (\mu_i^2) \wedge  X_i^{-2} [Ld\phi_i + A^i] \wedge  *_{(1,4)} F^i_{(2)} + \frac{L}{2} \sum_i \mu_i^2 X_i^{-2} F^i_{(2)} \wedge  *_{(1,4)} F^i_{(2)} \nonumber \\
& & -\frac{L}{2} \sum_i \mu_i^2 [Ld\phi_i + A^i] \wedge d(X_i^{-2} *_{(1,4)} F^i_{(2)} ) 
\end{eqnarray}
We can use the five-dimensional equations of motion~(\ref{eq:5deom1}) and~(\ref{eq:5deom2}) to replace the second and fifth terms. We find:
\begin{eqnarray}
   & & d \left(  \frac{L}{2} \sum_i \mu_i^2 \left[ X_i^{-1} *_{(1,4)} d X_i + X_i^{-2} [Ld\phi_i + A^i] \wedge *_{(1,4)} F^i_{(2)} \right] \right) \nonumber \\
&=&   \frac{L}{2} \sum_i d (\mu_i^2) \wedge X_i^{-1} *_{(1,4)} d X_i  + \frac{2}{L} \sum_{i j} \mu_i^2 M_{i j} \epsilon_{(1,4)} X_j^{-1} + \frac{L}{6} \sum_i X_i^{-2} F^i_{(2)} \wedge *_{(1,4)} F^i_{(2)} \nonumber \\
& & + \frac{L}{2} \sum_{i} d (\mu_i^2) \wedge  X_i^{-2} [Ld\phi_i + A^i] \wedge  *_{(1,4)} F^i_{(2)}  + \frac{L}{2} \sum_i \mu_i^2 [L d\phi_i + A^i] \bigwedge_{k \neq i} F^k_{(2)}
\end{eqnarray}
It is now easy to check, using the identity
$$ U = \sum_i(X_i^{-2}\mu_i^2 - \Delta X_i) = \sum_iX_i^{-1}(\mu_i^2 - 1) $$
that this differs from the expression for $G_{(5)}$ given in
equation~(\ref{5form}) by
$$ - \frac{4}{3L} \sum_i X_i^{-1} \epsilon_{(1,4)} -
    \frac{L}{2}\sum_i\mu_i^2[Ld\phi_i + A^i] \bigwedge_{k \ne i} F^k_{(2)}
 - \frac{L}{6} \sum_i X_i^{-2} F^i_{(2)} \wedge *_{(1,4)} F^i_{(2)} $$

We next evaluate
\begin{eqnarray}
 & & d \left( -\sum_i \left( \frac{X_i \mu_i}{\Delta} \right) L Z_i \bigwedge_k \mu_k [L d \phi_k + A^k]\right) \nonumber \\
 & = & -\sum_i \partial_{\nu} \left( \frac{X_i \mu_i}{\Delta} \right) dx_{\nu} L Z_i \bigwedge_k \mu_k [L d \phi_k + A^k] \nonumber \\
 & & - L \sum_i \left( \frac{X_i}{\Delta}(\mu_i^2 - 1) - \frac{X_i\mu_i}{\Delta^2}\sum_j 2\mu_jX_j(\mu_i\mu_j-\delta_{ij}) \right) W \bigwedge_k \mu_k [L d \phi_k + A^k] \nonumber \\
 & & -2LW \bigwedge_k \mu_k [L d \phi_k + A^k] + \left(\frac{1}{\Delta}\sum_i X_i - 3\right) LW \bigwedge_k \mu_k [L d \phi_k + A^k] \nonumber \\
 & & + \sum_{ij} \left( \frac{X_i \mu_i}{\Delta} \right) LZ_i \bigwedge_{k \ne j} \mu_k [L d \phi_k + A^k] \wedge \mu_jF^j_{(2)} \nonumber \\
 & = & -\sum_i \partial_{\nu} \left( \frac{X_i \mu_i}{\Delta} \right) dx_{\nu} L Z_i \bigwedge_k \mu_k [L d \phi_k + A^k] - \left(4 + \frac{2U}{\Delta^2} \right) L W \bigwedge_k \mu_k [L d \phi_k + A^k] \nonumber \\
 & & + \sum_{ij} \left( \frac{X_i \mu_i}{\Delta} \right) L Z_i \mu_j \bigwedge_{k \ne j} \mu_k [L d \phi_k + A^k] \wedge F^j_{(2)}
\end{eqnarray}

This expression can now be seen to differ from equation~(\ref{StarG5}) by
$$ 4LW\bigwedge_k \mu_k[Ld\phi_k + A^k] -
        L\sum_i Z_i \bigwedge_{k \ne i} \mu_k[Ld\phi_k + A^k] \wedge F^i_{(2)} $$
As well as the equations of motion~(\ref{eq:5deom2}),
we have used various useful identities which are listed below for convenience:
\begin{eqnarray}
dZ_i & = & 2\mu_i W \\
Z_i \wedge d\mu_j & = & (\delta_{ij} - \mu_i\mu_j)W \\
Z_{ij}\mu_k + Z_{ki}\mu_j + Z_{jk}\mu_i & = & 0 \\
Z_j - \frac{1}{\Delta}\sum_i X_i\mu_iZ_i\mu_j & = & -\frac{1}{\Delta}\sum_iX_i\mu_iZ_{ij} \label{Z_Delta}
\end{eqnarray}

So, adding together $G_{(5)}$ and $*G_{(5)}$, we see that
\begin{eqnarray}
F_{(5)} = & & d \left( \frac{L}{2} \sum_i \mu_i^2 \left[ X_i^{-1} *_{(1,4)} d X_i + X_i^{-2} [Ld\phi_i + A^i] \wedge  *_{(1,4)} F^i_{(2)} \right] \right) \nonumber \\
 & + & d \left( -\sum_i \left( \frac{X_i \mu_i}{\Delta} \right) L Z_i \bigwedge_k \mu_k [L d \phi_k + A^k]\right) \nonumber \\
 & - & \frac{4}{3L} \sum_i X_i^{-1} \epsilon_{(1,4)} - \frac{L}{6}\sum_i X_i^{-2} F^i_{(2)} \wedge  *_{(1,4)} F^i_{(2)} - 
    \frac{L}{2}\sum_i\mu_i^2[Ld\phi_i + A^i] \bigwedge_{k \ne i} F^k_{(2)} \nonumber \\
 & + & 4LW\bigwedge_k \mu_k[Ld\phi_k + A^k] -
        L\sum_i Z_i \bigwedge_{k \ne i} \mu_k[Ld\phi_k + A^k] \wedge F^i_{(2)}
\end{eqnarray}
The first two terms of this expression have been written as the derivative
of a globally well-defined potential. The next two terms are five forms on
$AdS_5$ and so are manifestly closed, but their integrals will depend on the
precise solution and are not needed here. The sum of the final three terms
is closed but not exact. We can (partly) integrate these terms locally for
$\mu_1 \neq 0$ to obtain the expression
$$ d \left( -\frac{L}{2}\mu_2^2 \bigwedge_{k \ne 3} [L d \phi_k + A^k] \wedge F^3_{(2)} +\frac{L}{2}\mu_3^2 \bigwedge_{k \ne 2} [L d \phi_k + A^k] \wedge F^2_{(2)}\right) $$
$$ + d \left( L\frac{Z_1}{\mu_1} \bigwedge_k \mu_k [L d \phi_k + A^k]\right) - \frac{L}{2} [Ld\phi_1 + A^1] \wedge F^2_{(2)} \wedge F^3_{(2)} $$
Putting this together with equation~(\ref{Z_Delta}) (for $j=1$) we can write the five-form
field strength in terms of a four-form potential, well-defined for
$\mu_1 \ne 0$, plus terms which do not couple to the D3-branes we
consider. So our final expression is
\begin{eqnarray}
 F_{(5)} = & & d \left( \frac{L}{2} \sum_i \mu_i^2 \left[ X_i^{-1} *_{(1,4)} d X_i + X_i^{-2} [Ld\phi_i + A^i] \wedge  *_{(1,4)} F^i_{(2)} \right] \right) \nonumber \\
 & + & d \left( -\frac{L}{2}\mu_2^2 \bigwedge_{k \ne 3} [L d \phi_k + A^k] \wedge F^3_{(2)} +\frac{L}{2}\mu_3^2 \bigwedge_{k \ne 2} [L d \phi_k + A^k] \wedge F^2_{(2)}\right) \nonumber \\
 & + & d \left( \frac{L}{\Delta \mu_1}\sum_i X_i \mu_i Z_{1i} \bigwedge_k \mu_k [L d \phi_k + A^k]\right) - \frac{L}{2} [Ld\phi_1 + A^1] \wedge F^2_{(2)} \wedge F^3_{(2)} \nonumber \\
 & - & \frac{4}{3L} \sum_i X_i^{-1} \epsilon_{(1,4)} - \frac{L}{6}\sum_i X_i^{-2} F^i_{(2)} \wedge  *_{(1,4)} F^i_{(2)}
\end{eqnarray}

\bibliographystyle{utphys}  
   
\bibliography{references}   
    
\end{document}